\documentclass[twocolumn,showpacs,preprintnumbers,amsmath,amssymb,aps, prl]{revtex4}

\usepackage{graphicx}
\usepackage{dcolumn}
\usepackage{bm}

\begin{document}

\title{New Method of Enhancing
Lepton Number Nonconservation
}

\author{M. Ikeda, I. Nakano, M. Sakuda, R. Tanaka, and M. Yoshimura}
\affiliation{%
Department of Physics, Okayama University \\
Tsushima-naka 3-1-1, Okayama 700-8530, Japan
}%

 \homepage{http://www.physics.okayama-u.ac.jp/}

\date{\today}

\begin{abstract}
The lepton number nonconserving (LENNON) conversion 
of type $ e^-  \rightarrow e^+$ in heavy atoms,
when irradiated by intense laser beam,
is considered to determine 
the Majorana nature and precise values of neutrino masses.
When the photon energy is fine tuned , 
the LENNON process is greatly enhanced by both the resonance effect
and a large occupancy of photons.
The signal of this process would be a positron of definite
energy, and a further nuclear $\gamma$-ray 
in a favorable case of transition to an excited nuclear level.
By constructing a united target-detector system, it is
possible to explore $O[1meV]$ range of the neutrino mass
parameter  with
a good positron energy resolution. 
\end{abstract}

\maketitle

Recent observations 
indicate that neutrinos have finite masses.
The immediate question that arises is whether these masses
are of Dirac or Majorana type. If they are of the Majorana type,
the lepton number is not conserved.
If this nonconservation were discovered, one should expect
great impacts on other areas of physics, 
including the possible explanation of
baryon-antibaryon imbalance in our universe \cite{th leptogenesis}.
The most extensively examined tool to study the Majorana nature of
the neutrino mass is the neutrinoless double beta
decay, and many proposals and experiments are already on-going
\cite{double beta limit}.

We wish to propose alternatives to the neutrinoless double beta
decay in order to investigate the nature of neutrino masses, 
both because the important issue of the lepton number nonconservation
(abbreviated as LENNON)
requires an independent experimental check, and because one should examine
it in many other processes.
We shall focus in the present work on photon stimulated 
electron conversion of type $ e^-  \rightarrow e^+$ 
in heavy atoms, with no accompanying neutrino.
The experimental signature of this process is
a monochromatic positron of few $MeV$, and furthermore
nuclear $\gamma$-ray in a favorable case of transition to nuclear
excited states.
Although simultaneous emission of low energy photons occurs, it might be
difficult to detect them, because their energies are
either down shifted or the same laser photon energy 
in the case of united target-detector
system later discussed, and they might easily be confused amidst the
background of high intensity beam.
We give rates both for this process and the background of
similar process of two accompanying neutrinos.

The basic low energy effective operator 
\cite{lty-04} that may arise in physics beyond the standard model is
\( \:
G_F^2 \tilde{m}_{\nu} ll \bar{q}q \bar{q}q
\: \)
where
$l$ is the lepton doublet and $q$ is the quark doublet having
the quantum number of the standard model.
This class of effective interaction violates the lepton number by two
units, $\Delta L = 2$, a class of simplest models predicting
the amplitude in direct proportion to some combination of
Majorana type of neutrino masses, $\tilde{m}_{\nu}$.
Thus, the rate of the non-radiative LENNON conversion of the type
$e^-(1s) \rightarrow e^+$, even for a favorable case of
$^{112}Sn$, becomes of order $10^{-29} year^{-1}$
for the Majorana neutrino mass of order $0.1 eV$.

\begin{figure}[htbp]
\begin{center}
\includegraphics[width=0.8\hsize,clip]{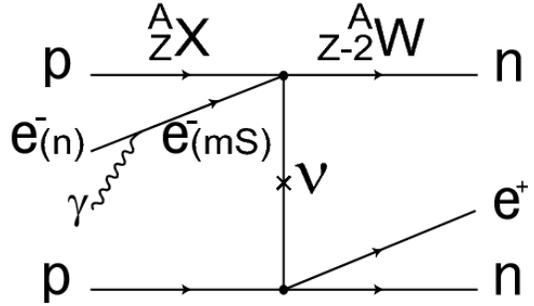}
\caption{Feynman diagram for photon irradiated LENNON process of
atomic transition $n \rightarrow ms$.}
\label{fig:fig1}
\end{center}
\end{figure}

\begin{figure}[htbp]
\begin{center}
\includegraphics[width=0.8\hsize,clip]{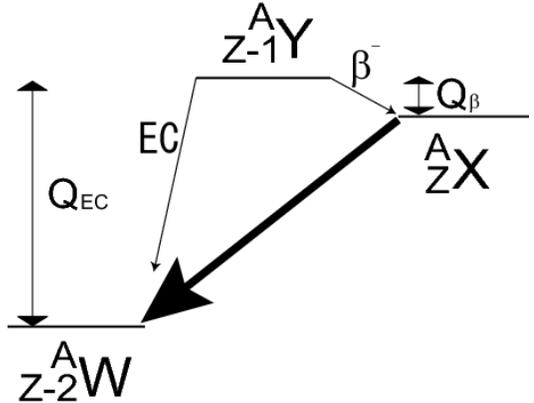}
\caption{Level structure for three adjacent nuclei of
mass number $A$ and atomic number $Z$.}
\label{fig:fig2}
\end{center}
\end{figure} 

We instead consider LENNON conversion
stimulated by radiative absorption, of the type depicted in Fig. 1.
Here $^{A}_{Z} X$ is a nucleus of mass number $A$ and atomic number $Z$.
An electron occupying an atomic $n$ state is upshifted by
photon irradiation to an unoccupied $ms$ level, 
and later captured by nucleus, producing a positron
by $e^- \rightarrow e^+$ conversion.
Another possibility is the positron emission first by weak interaction
and a subsequent absorption of photon by atomic electron of $^{A}_{Z -1} Y$.
The idea is to enhance the $e^- \rightarrow e^+$ conversion with the help
of both the resonance effect and a high occupancy of
laser photons in a quantum level.
Without irradiation of the initial photon,
this is a process conjugate to the neutrinoless double
beta decay, and can explore the same combination of 
the neutrino mass parameter $m_{ee}$, where
$m_{\alpha \beta} = \sum_k U_{\alpha k}U_{\beta k}m_k$
with $m_k$ 3 neutrino mass eigenvalues.

In Fig.2 we depict relevant nuclear levels for three adjacent nuclei
of the same mass number.
The final nucleus $^{A}_{Z-2} W$ can be in an excited nuclear level, for which
triple coincidence for detection including nuclear $\gamma$-ray
may become possible.
If both $\beta^-$ decay and electron capture rates 
of the intermediate $^A_{Z-1} Y$ nucleus are experimentally
known, one may reliably estimate the rate for the radiative 
$e^- \rightarrow e^+$ conversion.
The best candidate with this regard
is the isotope $^{112}_{50}Sn$ \cite{Sn}.

A straightforward computation for the atomic
$n \rightarrow m$ transition followed by the capture
gives the following matrix element;
\begin{widetext}
\begin{multline}
\frac{e G_F^2 m_{ee}}{4\pi }\,2\pi \delta(E_{\gamma} + \Delta E_{i,f}
- E_+  + \Delta \epsilon_{nm}) \\
\times \int d^3 x d^3 y
\langle f| \frac{J(\vec{x})\cdot J(\vec{y})}{|\vec{x} - \vec{y}|} 
| i \rangle
\langle \vec{p}_+|\overline{e^c}(\vec{x})(1 - \gamma_5) e(\vec{y})
\sum_m \left(  | m \rangle
\frac{\langle m|H_{\gamma}| n\rangle }
{E_{\gamma} - \Delta \epsilon_{nm} + i\tilde{\Gamma}_{n}/2}  \right)\, ,
\label{matrix el}
\end{multline}
\end{widetext}
where $e(\vec{y})$ etc. is the electron field operator.
Here $\Delta E_{i,f} = E_i - E_f $, with energies $E_{i \,, f}$ 
referring to nuclear levels of initial and
final states. Thus,
$E_+ = \Delta E_{i,f}+ E_{\gamma}  + \Delta \epsilon_{nm}$ is
the monochromatic positron energy.
The important natural width $\tilde{\Gamma}_{n}$ in (\ref{matrix el})
refers to that of the  hole of $n$ atomic state.
The energy difference $\Delta \epsilon_{nm} = \epsilon_n - \epsilon_m$
where $\epsilon_{n\,,m}$ are energies of atomic levels.

The neutrino propagator has been replaced by
the instantaneous Coulomb potential, which is allowed
since for low energy electrons the energy transfer is small;
$|q_0| \ll |\vec{q}|$. 
At low energies the electron wave functions can be taken out from
the integral (\ref{matrix el}), and one may
separate out the nuclear matrix element. Furthermore,
by introducing an average inter-proton distance,
$R_n$, one may take out the factor 
$\langle 1/|\vec{x} - \vec{y}| \rangle = 1/R_n$
outside the nuclear matrix element.
We use 
\(\:
R_n \approx (0.82 A^{1/3} + 0.58) fm
\,,
\:\)
following \cite{Sn}.

The radiative electron conversion via $ms$ atomic state
has a cross section of the form,
\begin{eqnarray}
\Gamma_{0\nu}^{(mS)}\,\frac{|\langle mS| H_{\gamma}
| n, \gamma \rangle|^2}{(E_{\gamma} - \Delta \epsilon_{nm})^2 +
\tilde{\Gamma}_{n}^2/4}
\,,
\label{radiative rate}
\end{eqnarray}
where $H_{\gamma} $ is QED interaction.
Let us first consider
the non-radiative conversion rate $\Gamma_{0\nu}^{(ms)} 
= |\psi_{ms}(0)|^2 \sigma_{0\nu}$. 
This product is an effective luminosity $|\psi_{ms}(0)|^2$
of confined $ms$ electron times the cross section for
$e^- ({\rm free}) +  ^{A}_{Z} X  \rightarrow e^+ + ^{A}_{Z-2} W$ given by
\begin{eqnarray}
\sigma_{0\nu} =
\frac{G_F^4 \tilde{m}_{\pi}^2 |m_{ee}|^2}{16\pi^3} p_+ E_+ \,, \nonumber\\
\tilde{m}_{\pi}
\equiv
\int d^3 x d^3 y
\frac{\langle f | J(\vec{x})\cdot J(\vec{y})| i \rangle }
{|\vec{x} - \vec{y}|}
\,.
\end{eqnarray}
Typically, $\sigma_{0\nu} = O[10^{-66} cm^{-2}] |m_{ee}/1 eV|^2 
(p_+ E_+/MeV^2)$.
When a photon is irradiated, the product is further multiplied
by the Breit-Wigner function, which can be very large near
the resonance, $E_{\gamma} = \Delta \epsilon_{nm} \equiv E_0$.
In other words, the cross section ratio of the photon-irradiated conversion, 
(\ref{radiative rate}) relative to the elementary $\sigma_{0\nu}$, is
\begin{eqnarray}
K(E) &=&
\frac{|\langle mS| H_{\gamma}
| n, \gamma \rangle|^2|\psi_{ms}(0)|^2}{(E - E_0)^2 +
\tilde{\Gamma}_{n}^2/4} \nonumber\\
&\approx& \frac{2\pi^2 \Gamma_{mS \rightarrow n}|\psi_{ms}(0)|^2}
{E_0^2\,[(E - E_0)^2 + \tilde{\Gamma}_{n}^2/4]}
\,.
\label{flux modified}
\end{eqnarray}
Here the branching fraction
$B_{ms \rightarrow n} =\Gamma_{ms \rightarrow n}/\tilde{\Gamma}_{n}$ 
is of order unity.

The actual rate, when the photon beam of luminosity density $I(E)$,
within unit energy bin and per unit area times unit time,
is irradiated, is given by
\begin{eqnarray}
R &=& \int dE \, I(E) \sigma_{\gamma e^- \rightarrow e^+} \nonumber\\
  &=& S\,\int dE \, \frac{I(E)}{(E - E_0)^2 + \tilde{\Gamma}_{n}^2/4}
\,.
\label{resonance integral}
\end{eqnarray}
Note the dimensionless strength factor given by
\begin{eqnarray}
S &=& 2\pi^2 \sigma_{0\nu}|\psi_{ms}(0)|^2
\frac{B_{mS \rightarrow n}\tilde{\Gamma}_{n}}{E_0^2}\nonumber\\
&\approx&
2\pi \sigma_{0\nu}(Z\alpha)^6 m_e^3
\frac{B_{ms \rightarrow n}\tilde{\Gamma}_{n}}{E_0^2}
\,.
\end{eqnarray}
The integral in (\ref{resonance integral}) may readily be
evaluated when the photon flux has little variation over the natural
width $\tilde{\Gamma}_{n}$ around $E_0$:
\begin{eqnarray*}
\int dE \frac{I(E)}{(E - E_0)^2 + \tilde{\Gamma}_{n}^2/4}
\approx 2\pi \frac{I(E_0)}{\tilde{\Gamma}_{n}}
\,.
\end{eqnarray*}
When the photon beam is tuned and the resonance $E_0$ is not
missed, one further has
\( \:
I(E_0) \approx F(E_0)/\Delta E
\,,
\: \)
where $\Delta E$ is the energy resolution of photon beam and 
$F(E_0)= \int dE I(E)$ is the total flux integrated over the region 
around $E_0$.
This leads to the rate formula,
\begin{equation}
R \approx \frac{2\pi^2 F(E_0)}{E_0^2 \Delta E}B_{ms \rightarrow n}
2\pi |\psi_{ms}(0)|^2\sigma_{0\nu}
\,.
\label{rate formula}
\end{equation}

Each factor has a clear meaning; the first 
$F(E_0)/(E_0^2 \Delta E(2\pi^2 )^{-1}\,)$ expressing
the number of occupied photons
within the relevant quantum phase space,
the second $B_{ms \rightarrow n}$ the branching fraction
of order unity,
the last $|\psi_{ms}(0)|^2\sigma_{0\nu}$ the $e^-
\rightarrow e^+$ conversion rate from $ms$ state.
Thus, the enhancement or reduction factor by photon irradiation
relative to the capture rate $|\psi_{1s}(0)|^2\sigma_{0\nu}$
is given by
\( \:
4\pi^3 F(E_0)\,(|\psi_{ms}(0)|^2/E_0^3)\,(E_0/\Delta E)
B_{ms \rightarrow n}/m^6
\,.
\: \)
It is useful to define a  
quality factor $Q$ of photon beam defined by
$Q = 4\pi^3 F(E_0) /(E^2_0 \Delta E)$, which is numerically
$2 \times 10^{10} (F(E_0)/1 kW mm^{-2})(E_0 / eV)^{-4}(10^{-9}E_0 /\Delta E)$.
An advantage of a strong laser beam, which can give a large $Q$, 
is obvious, when compared to X-ray which typically gives
$Q \ll 1$ (except contemplated X-ray laser).

One might wonder the validity of the linear rise with the laser power
of the rate, eq.(\ref{rate formula}), because irradiated atoms may
become transparant once electrons in the lower energy level
are completely lifted to the higher energy level.
However, we are considering the situation of equilibrated atoms
between the two levels by constant irradiation of laser beam.
In this case the rate formula (\ref{rate formula}) is still valid,
with a minor multiplication of the population factor in the lower level.
The formula valid in the large power limit is different and
shall be presented elsewhere by one of the present authors (MY).

We shall next discuss how to estimate
the rate $\Gamma_{0\nu}$ for the non-radiative electron conversion by nucleus
that appears in the fundamental formula (\ref{radiative rate}).
By truncating the nuclear level sum 
in $\sum_k \langle f| J|k \rangle \langle k| J| i\rangle $ 
by a single ground state $|k \rangle = |  ^{A}_{Z-1} Y \,,1^+ \rangle$,
one may replace the nuclear matrix element
$\langle k| J | i \rangle$ by the beta decay rate, 
and $\langle f| J| k \rangle $ by the electron capture rate
of nucleus $^A_{Z-1} Y$.
In the case of the nonradiative decay, $e^- (1s) \rightarrow e^+$ process,
the rate becomes
\begin{eqnarray}
&&
\Gamma_{0\nu} = \frac{3\pi^3|m_{ee}|^2}{4R_n^2}
\frac{p_+ E_+ }{p_{\nu}^2 \Delta_{\beta}^5 I}\Gamma_{EC}\Gamma_{\beta}
\,,
\label{rate for 0n}
\end{eqnarray}
where $\Gamma_{EC}\,, \Gamma_{\beta}$ are the electron capture and
the beta decay rate of $^A_{Z-1} Y$, and
$p_i$ are respective lepton momenta.
The maximum $\beta$ energy $\Delta_{\beta} = Q_{\beta} + m_e
- \epsilon_{1s}$, and $p_{\nu} = Q_{EC} - \epsilon_{1s}$.
Here $I$ is a dimensionless phase space factor for the beta decay
and $1/30$ in the limit of zero electron mass.
We ignored the difference of $1s$ electron wave function of
$^A_{Z} X$ and $^A_{Z-2} W$,
whose error should be small, of order $4/Z$.

For instance, in the case of $0^+ \rightarrow 0^+$ nuclear
transition of $^{112}_{50}Sn \rightarrow ^{112}_{48}Cd$,
both intermediate $^{112}_{49}In$ (of $1^+$) $\beta^-$ decay and electron 
capture rates are known \cite{nuclear data}.
Thus, the rate computed according to eq.(\ref{rate for 0n}) is 
$\approx (2\times 10^{29} y )^{-1} (|m_{ee}|/0.1 eV)^{-2}$ 
for $^{112}_{50}Sn$.
It appears that nuclear matrix elements are large,
as pointed out in \cite{Sn}.

We numerically give the laser irradiated
$e^- \rightarrow e^+$ conversion rate for one nucleus target;
\begin{eqnarray}
R &\approx & 5\times 10^{-35} y^{-1}\nonumber\\
&&\times \frac{\Gamma_{0\nu}}{10^{-29} y^{-1}}\,
\left( \frac{4}{m} \right)^{6}\frac{E_{\gamma}}{\Delta E}
\frac{F}{W mm^{-2}}\left( \frac{E_{\gamma}}{eV} \right)^{-4}\,.
\label{flip rate}
\end{eqnarray}
The crucial factor to obtain a large enhancement
of the rate is the inverse of resolution, and 
with $\Delta E/E_{\gamma} = 10^{-9}$ available commercially, the
rate is enhanced by $\approx 5 \times 10^{6}\,(4/m)^{6}$, 
using a laser beam power $1 kW/mm^2$.
Linear increase with the intensity $F$ 
of this event rate is the key check point
of experimental verification of the process.
A few isotopes which may give rates of $\Gamma_{0\nu} > 10^{-31} y^{-1}$
are illustrated in Table 1.

\begin{table}
\caption{\label{tab:isotopes} Useful characteristics for a few isotopes} 
\begin{ruledtabular}
\begin{tabular}{l|cccc}

Atom &  abnd.\% & $Q_{e^{+}}$ (keV)&$\tau_{0\nu \frac{1}{2}}$yr $|\frac{m_{ee}}{0.1eV}|^{-2}$
&$\frac{\Gamma_{0\nu}}{\Gamma_{2\nu}(\delta E)}\footnotemark[1]$ \\[4pt] 
\hline\\[-5pt]
$^{78}_{36}$Kr & 0.35 & 1846  & $1.5\times 10^{29} \footnotemark[2]$  & $1.6\times 10^{3}$ \\[3pt] 
$^{112}_{ 50}$Sn  & 0.97 &  901 & $5.5\times10^{28}$& $1.8\times 10^{3}$ \\[3pt]
$^{130}_{56}$Ba  & 0.11 &1588 & $2.3\times10^{29}$ & $1.7\times 10^{3}$ \\[3pt]

\end{tabular}
\end{ruledtabular}
\footnotetext[1]{ We assume $0.1 eV$ of the neutrino mass, and $\delta E = 100 keV$. }
\footnotetext[2]{For $Kr$ half-life, we follow the calculation of \cite{Sn} 
intead of eq.(\ref{rate for 0n}).}
\end{table} 

The required resonance tuning might be a great practical obstacle 
since laser frequencies are superposition of
quantized level differences.
(Alternatively, the use of laser with a continuous spectrum might help greatly.)
It is however possible to avoid this problem by selecting the same
lasing medium as the target nucleus, thus we arrive at
the concept of a united target-detector system.
A good target must then be both lasing and an efficient 
$e^- \rightarrow e^+$ converter.
Candidates of such targets are limited, but the element $Kr$ is a good example.
The $Kr^+$ laser uses the inverted population of
$4p$ atomic state, which falls down to $4s$ by stimulated
emission caused by irradiated laser \cite{laser}. 
In this basic process $4s$ electron may very rarely be captured 
by $0^+$ $Kr$ nucleus to emit a positron of energy $\approx 1.8 MeV$.

The isotope $^{78}_{36}Kr$ after the $e^- \rightarrow e^+$ conversion 
ends up with $^{78}_{34}Se$, which has a few
excited energy levels of $0^+$ below the $^{78}_{36}Kr$ ground level.
Thus, nuclear gamma rays are expected to give an opportunity of
coincident measuremet.

Let us estimate what might occur within an ideal $Kr^+$ 
or gaseous laser device containing $Kr$ such as $KrF$ excimer laser.
Suppose that the gas chamber of the device contains $10^4  \, cm^3$ of 1 ATM 
enriched $ ^{78}_{36}Kr$,
which has $\approx 3 \times 10^{23}$ $^{78}_{36}Kr$ atoms.
The formula (\ref{flip rate}) tells that $e^- \rightarrow e^+$ conversion
occurs with a rate,
\( \:
\approx 10^4 /y \,(I/MW \,mm^{-2}) \,,
\: \)
assuming a resolution $\Delta E/E_{\gamma} = 10^{-9}$  
and $\Gamma_{0\nu} = 10^{-29}/y$.
For the $Kr^+$ laser, $m = 4 \,, E_{\gamma} = 1.9 eV$ 
for one of the main lines.
The rate scales with the neutrino mass parameter as
$\Gamma_{0\nu} = 0.9 \times 10^{-29} y^{-1}|m_{ee}/0.1 eV|^2$,
using nuclear matrix elements of \cite{Sn} adopted to $^{78}_{36}Kr$, thus
one may be able to explore $|m_{ee}|$ down to $1 meV$.

The important physics background to the present process
is the corresponding process of two accompanying neutrinos
caused by the second order weak interaction,
which itself is of interest.
We may estimate this rate by using the same approximation of
one-level truncation.
The result is given by the ratio of two processes,
\( \:
\Gamma_{0\nu}/\Gamma_{2\nu} \approx (20160\pi^2 |m_{ee}|^2 m_{\pi}^2)/
(\Delta^6 R_n^2)
\: \)
which is $\approx 0.66 \times
10^{-4}|m_{ee}/1 meV|^2 (m_{\pi}R_n)^{-2}(\Delta/MeV)^{-6}$.

On the other hand, the ratio when the positron energy
is limited near the end point of $e^+$ energy width $\delta E$ is 
\( \:
\Gamma_{0\nu}/\Gamma_{2\nu}(\delta E) \approx
45\pi^2 m_{\pi}^2 |m_{ee}|^2/(R_n^2 (\delta E)^6)
\,.
\: \)
With $\delta E \approx 100 keV$ and for $^{78}_{36}Kr$,
\begin{equation}
\frac{\Gamma_{0\nu}}{\Gamma_{2\nu}(\delta E)} \approx
O[18] \left| \frac{m_{ee}}{0.01 eV} \right|^{2}\left( \frac{\delta E}{100 keV} \right)^{-6}
\,.
\end{equation}
The positron energy spectrum for two neutrino process is given by
\begin{eqnarray}
d \Gamma_{2\nu} \propto E \sqrt{E^2 - m_e^2} (E - \Delta)^5 dE
\,.
\end{eqnarray}
The signal of LENNON would be an excess of positrons near the end point of
two neutrino processes.
In the favorable case to decay into a nuclear excited level the background
$(2 \nu)$ process is suppressed by the phase space
factor $(E - \Delta)^5 $ if the excited level has a relatively
high energy.

\vspace{0.5cm}
In summary, it appears possible to explore the Majorana neutrino mass range
down to $1 meV$ by laser irradiated $e^- \rightarrow e^+$ conversion, when
a lasing gas medium $> 10^4\, cm^3$
of high power $>1 MW$ or more, is used,
along with a good measured positron energy resolution of $< 50 keV$.

\end{document}